\pdfoutput=1

\documentclass[sigconf,natbib=true,anonymous=false]{acmart}

\setcopyright{none}
\renewcommand\footnotetextcopyrightpermission[1]{}

\usepackage[T1]{fontenc}

\usepackage[hang,flushmargin]{footmisc}

\usepackage{url}            
\usepackage{booktabs}       
\usepackage{amsfonts}       
\usepackage{nicefrac}       
\usepackage{microtype}      
\usepackage{amsmath}
\usepackage{enumitem}
\usepackage{graphicx}       
\usepackage{caption}
\usepackage{subcaption}
\usepackage{pifont}
\usepackage{multirow}

\newcommand{\cmark}{\ding{51}}%
\newcommand{\xmark}{\ding{55}}%

\newcommand\ignore[1]{}
\newcommand\red[1]{\textcolor{red}{#1}}

\title{Simple Yet Effective Neural Ranking and Reranking Baselines for Cross-Lingual Information Retrieval}

\author{Jimmy Lin}
\affiliation{University of Waterloo \country{Canada}}

\author{David Alfonso-Hermelo}
\affiliation{Huawei Noah's Ark Lab \country{Canada}}

\author{Vitor Jeronymo}
\affiliation{UNICAMP \country{Brazil}}

\author{Ehsan Kamalloo}
\affiliation{University of Waterloo \country{Canada}}

\author{Carlos Lassance}
\affiliation{Naver Labs Europe \country{France}}

\author{Rodrigo Nogueira}
\affiliation{UNICAMP \country{Brazil}}

\author{Odunayo Ogundepo}
\affiliation{University of Waterloo \country{Canada}}

\author{Mehdi Rezagholizadeh}
\affiliation{Huawei Noah's Ark Lab \country{Canada}}

\author{Nandan Thakur}
\affiliation{University of Waterloo \country{Canada}}

\author{Jheng-Hong Yang}
\affiliation{University of Waterloo \country{Canada}}

\author{Xinyu Zhang}
\affiliation{University of Waterloo \country{Canada}}

\begin{document}

\renewcommand{\shortauthors}{}
\pagestyle{empty}

\begin{abstract}
The advent of multilingual language models has generated a resurgence of interest in cross-lingual information retrieval (CLIR), which is the task of searching documents in one language with queries from another.
However, the rapid pace of progress has led to a confusing panoply of methods and reproducibility has lagged behind the state of the art.
In this context, our work makes two important contributions:
First, we provide a conceptual framework for organizing different approaches to cross-lingual retrieval using multi-stage architectures for mono-lingual retrieval as a scaffold.
Second, we implement simple yet effective reproducible baselines in the Anserini and Pyserini IR toolkits for test collections from the TREC 2022 NeuCLIR Track, in Persian, Russian, and Chinese.
Our efforts are built on a collaboration of the two teams that submitted the most effective runs to the TREC evaluation.
These contributions provide a firm foundation for future advances.
\end{abstract}
\maketitle

\section{Introduction}

Cross-lingual information retrieval (CLIR) is the task of searching documents in one language with queries from a different language---for example, retrieving Russian documents using English queries.
Typically, a CLIR system exists as part of an overall pipeline involving machine translation, related human language technologies, and sometimes human experts, that together help users satisfy information needs with content in languages they may not be able to read.
Research on cross-lingual information retrieval dates back many decades~\cite{Hull_Grefenstette_SIGIR1996,Federico_Bertoldi_SIGIR2002,WangJ06,Nie_2010}, but there has been a recent revival of interest in this challenge~\cite{Yu_Allan_SIGIR2020,Galuscakova:2111.05988:2021}, primarily due to the advent of multilingual pretrained transformer models such as mBERT~\cite{bert} and XLM-R~\cite{xlmr}.

A nexus of recent research activity for cross-lingual information retrieval is the TREC NeuCLIR Track, which ran for the first time at TREC 2022 but has plans for continuing in 2023 and perhaps beyond.
The track provides a forum for a community-wide evaluation of CLIR systems in the context of modern collections and systems, dominated today by neural methods.
NeuCLIR topics (i.e., information needs) are expressed in English, and systems are tasked with retrieving relevant documents from corpora in Chinese, Persian, and Russian.

Perhaps as a side effect of the breakneck pace at which the field is advancing, we feel that there remains a lack of clarity in the IR community about the relationship between different retrieval methods (e.g., dense vs.\ sparse representations, ``learned'' vs.\ ``heuristic'' vs.\ ``unsupervised'', etc.) and how they should be applied in different retrieval settings.
Furthermore, the increasing sophistication of today's retrieval models and the growing complexity of modern software stacks create serious challenges for reproducibility efforts.
This not only makes it difficult for researchers and practitioners to compare alternative approaches in a fair manner, but also creates barriers to entry for newcomers.
These issues already exist for mono-lingual retrieval, where documents and queries are in the same language.
With the added complexity of cross-lingual demands, the design choices multiply (choice of models, training regimes, application of translation systems, etc.), further muddling conceptual clarity and experimental reproducibility.

\paragraph{Contributions}
Our work tackles these challenges, specifically focused on helping both researchers and practitioners sort through the panoply of CLIR methods in the context of modern neural retrieval techniques dominated by deep learning.
Our contributions can be divided into a ``conceptual'' and a ``practical'' component:

Conceptually, we provide a framework for organizing different approaches to cross-lingual retrieval based on the general design of multi-stage ranking for mono-lingual retrieval.
These architectures comprise first-stage retrievers that directly perform top-$k$ retrieval over an arbitrarily large collection of documents, followed by one or more reranking stages that refine the rank order of candidates generated by the first stage.

Recently, Lin~\cite{Lin_SIGIRForum2021} proposed that retrieval techniques can be characterized by the representations that they manipulate---whether dense semantic vectors or sparse lexical vectors---and how the weights are assigned---whether heuristically, as in the case of BM25, or by a neural network that has been trained with labeled data.
Translated into the cross-lingual case, this leads naturally to three main approaches to first-stage retrieval:\ document translation, query translation, and use of language-independent representations.
While these approaches date back many decades, there are ``modern twists'' based on {\it learned} representations that take advantage of powerful pretrained transformer models.

\begin{figure*}[t]
\centering
\includegraphics[width=0.95\textwidth]{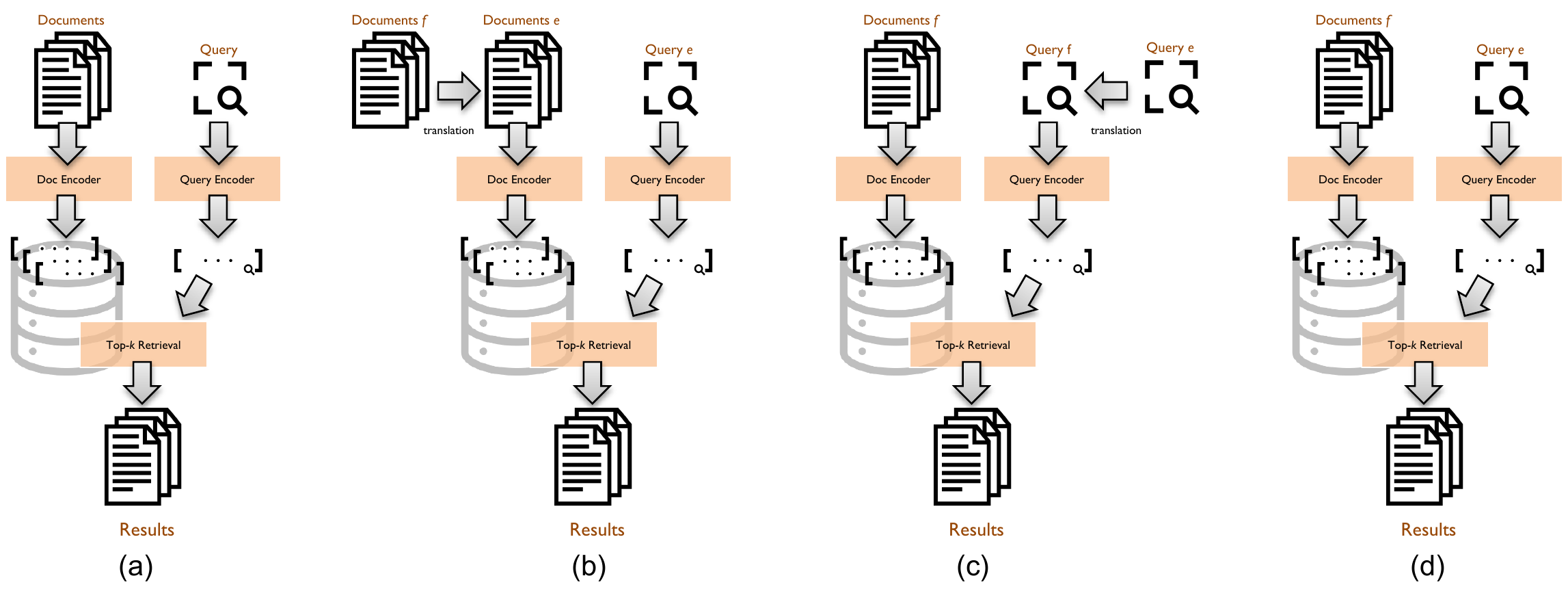}
\caption{Different retrieval architectures: (a) a mono-lingual bi-encoder architecture that captures both dense and sparse retrieval methods; (b) bi-encoder adapted for document translation, where all documents are translated into $e$ and queries remain in $e$; (c) bi-encoder adapted for query translation, where query $e$ is translated into $f$ and issued against documents in $f$; (d) bi-encoder where the encoders can project content from multiple languages into the same representation space.}
\label{fig:architectures}
\end{figure*}

For mono-lingual retrieval, a standard multi-stage architecture applies rerankers to the output of first-stage retrievers, like those discussed above.
In a cross-lingual context, we describe how {\it cross-lingual} rerankers can be designed and built using existing multilingual models.
Results fusion forms the final component of our conceptual framework.
Within a multi-stage architecture, there arises a natural question of {\it when} fusion should be performed:\ this manifests in the early vs.\ late fusion techniques that we examine.

Practically, we provide a number of reproducible baselines in the context of the above conceptual framework for the TREC 2022 NeuCLIR test collection, including variants of the highest-scoring runs that were submitted to the evaluation.
These reproducible baselines have been incorporated into the Anserini and Pyserini IR toolkits.
Our efforts are built on a collaboration of the two teams that submitted the most effective runs to the TREC evaluation.

We hope that this work provides a solid foundation for future work, both in terms of offering a conceptual framework and reference implementations that the community can further build on.

\section{Mono-Lingual Retrieval Overview}
\label{section:background}

Since mono-lingual retrieval architectures provide the starting point for cross-lingual retrieval, it makes sense to begin with an overview of modern mono-lingual methods.
Here, we adopt the standard formulation of the (mono-lingual) retrieval task (also called {\it ad hoc} retrieval).
From a finite but arbitrarily large collection of documents $\mathcal{C} = \{d_1, d_2 \ldots, d_n\}$, the system's task, given query $q$, is to return a top-$k$ ranking of documents that maximizes some metric of quality such as nDCG or average precision.

\paragraph{Rerankers}
The earliest applications of neural networks to tackle {\it ad hoc} retrieval in a data-driven manner date back to the mid 2000s in the context of learning to rank~\cite{Burges_etal_ICML2005}.
Since then, search engine design has been dominated by multi-stage ranking architectures~\cite{Matveeva_etal_SIGIR2006,Wang_etal_SIGIR2010}, where a first-stage retriever (often, just BM25 retrieval) generates candidate documents that are then reranked by one or more stages, typically by machine-learned models.
In the ``transformer era'', for example, BERT~\cite{Nogueira:1901.04085:2019,nogueira2019multi} and T5~\cite{nogueira2020document} can be used in exactly this manner.
Use of pretrained transformers for reranking requires feeding the model both the query and the candidate text, and this style of model application is known as a cross-encoder.

\paragraph{Bi-encoder architectures}
An important recent innovation for passage retrieval was the introduction of so-called dense retrieval models that take advantage of a bi-encoder design (contrasted with the cross-encoder design discussed above): DPR~\cite{dpr} and ANCE~\cite{ance} are two early examples.
With sufficient labeled data, we can learn encoders (typically, transformer-based models) that project queries and documents into a dense (semantic) representation space (e.g., 768 dimensions) where relevance ranking can be recast as nearest-neighbor search over representation vectors.

After the introduction of dense retrieval models, researchers soon realized that transformer-based encoders could also be coaxed to generate {\it sparse} representations, where the vector basis, for example, spans the input vocabulary space.
Another way to view these so-called sparse retrieval models is to contrast them with BM25:\
whereas BM25 term weights are assigned using a heuristic scoring function, sparse retrieval models assign term weights that are {\it learned} using pretrained transformers such as BERT.
Examples of these learned sparse retrieval models include DeepImpact~\cite{mallia2021learning}, uniCOIL~\cite{lin2021few,zhuang2021fast}, SPLADE~\cite{splade}, as well as many others.

Recently, Lin~\cite{Lin_SIGIRForum2021} made the observation that dense retrieval models, sparse retrieval models, and traditional bag-of-words models (e.g., BM25) are all parametric variations of a bi-encoder architecture, which is shown in Figure~\ref{fig:architectures}(a).
In all three classes of models, ``encoders'' take queries or documents and generate vector representations.
There are two major axes of differences, the first of which lies in the basis of the representation vector:\ dense retrieval models generate dense (semantic) representations whereas sparse retrieval models and bag-of-words model ground their representation vectors in lexical space.
The other major axis of variation is whether these representations are learned:\ {\it yes} in the case of dense and sparse retrieval models, but {\it no} in the case of traditional bag-of-words models.
The conceptual framework for mono-lingual retrieval provides us with a basis for organizing cross-lingual retrieval approaches, which we discuss next.

\section{Cross-Lingual Retrieval Methods}
\label{section:clir}

The cross-lingual information retrieval task is formalized in a similar manner as the mono-lingual retrieval task.
We assume a collection of documents $\mathcal{C}_f$ in language $f$ comprised of $\{d_1, d_2 \ldots, d_n\}$.
The system is given a query $q$ in language $e$, which we denote $q_e$ for clarity, and its task is to return a top-$k$ ranking of documents from $\mathcal{C}_f$ that maximizes some metric of quality such as nDCG or average precision.
Throughout this work, $e$ refers to English and $f$ refers to some non-English language (e.g., Russian), but this need not be the case in general.

Building from the design of the mono-lingual retrieval architecture presented in the previous section, our discussions begin with three possible designs for first-stage retrieval:\ document translation, query translation, and the use of language-independent representations.
We then overview cross-encoders for reranking the output of first-stage retrievers and finally conclude with some thoughts about fusion techniques.

To further ground cross-lingual retrieval techniques, we provide some details about the TREC 2022 NeuCLIR evaluation.
Given English queries, participants are tasked with retrieving from three separate corpora comprising Persian, Russian, and Chinese newswire documents curated from the Common Crawl between August 1, 2016 and July 31, 2021.
The corpora are modest in size, with 2.23 million documents in Persian, 4.63 million documents in Russian, and 3.18 million documents in Chinese.

Information needs (i.e., topics, in TREC parlance) were developed following a standard process for building retrieval test collections~\cite{Voorhees_CLEF2002,Harman_2011}.
The organizers released 114 topics, originally developed in English, which were then translated into Persian, Russian, and Chinese---both by humans and automatically by Google Translate.
The topics comprise ``title'' and ``description'' fields, where the former are akin to keyword queries and the latter are roughly sentence-long articulations of the information need.
By design, all topics are aligned, in the sense that for each topic, we have translations in all three languages.
However, it was {\it not} the case that all topics were evaluated for all languages:\ 
In total, the organizers released relevance judgments for 46 topics in Persian, 45 topics in Russian, and 49 topics in Chinese.

\subsection{Document Translation}

A very simple approach to cross-lingual information retrieval is known as document translation:\ 
Given $q_e$ in language $e$ and the corpus $\mathcal{C}_f$ in language $f$, we can translate the entire corpus into language $e$, i.e., $\{ \textsc{Translate}(d_i) \}$, and then perform mono-lingual retrieval in language $e$.
This design is shown in Figure~\ref{fig:architectures}(b), where the primary addition is a document translation phase that feeds into the document side of the bi-encoder architecture.

While translating the entire corpus can be time-consuming, it only needs to be performed once and can be viewed as an expensive pre-processing step, like other computationally demanding document expansion techniques such as doc2query~\cite{Nogueira_etal_arXiv2019}.
Any translation technique can be used, including off-the-shelf MT systems.
Generally, since documents are comprised of well-formed sentences, automatic translation output can be quite fluent, depending on the quality of the underlying system.
This stands in contrast to query translation (see below), where quality often suffers because queries are usually much shorter (hence lacking context) and systems are not usually trained on such inputs.

Once $\mathcal{C}_f$ has been translated into $\mathcal{C}_e$, we now have a mono-lingual retrieval task since queries are also in $e$.
In our case, the three corpora are in Persian, Russian, and Chinese, and we used the English translations provided by the NeuCLIR Track organizers, generated by the SockEye MT system.
From the NeuCLIR topics, we extracted three types of English queries:\ only  the ``title'' field, only the ``description'' field, and both.
Our experiments used two retrieval models and pseudo-relevance feedback:

\paragraph{BM25}
Despite the advent of numerous neural ranking models, this traditional ``bag-of-words'' model remains a robust baseline.

\paragraph{SPLADE}
We chose SPLADE++ Ensemble Distil~\cite{splade} due to its zero-shot capabilities.
The SPLADE family of models is a sparse neural retrieval model that learns both document and query expansion controlled by a regularization term.

\paragraph{Pseudo-relevance feedback (PRF)}
On top of results from both BM25 and SPLADE, we apply pseudo-relevance feedback.
While RM3 is a popular choice and has been well studied in the context of neural methods~\cite{Yang_etal_SIGIR2019}, in this work we instead apply Rocchio feedback, for two reasons:\ First, Rocchio feedback has been demonstrated to be an effective pseudo-relevance feedback approach for dense vector representations, and applying Rocchio to lexical representations provides conceptual unity.
In contrast, there is no equivalent RM3 variant for dense vectors, which makes comparing sparse and dense PRF more difficult.
Second, previous work has shown that Rocchio is at least as effective as RM3~\cite{Yuqi}, so we gain simplicity and consistency without sacrificing effectiveness.

\subsection{Query Translation}

The flip side of document translation is known as query translation:\
Given $q_e$ in language $e$ and the corpus $\mathcal{C}_f$ in language $f$, we can translate the query into language $f$, i.e., $\textsc{Translate}(q_e) = q_f$, and then perform mono-lingual retrieval in language $f$.
This design is shown in Figure~\ref{fig:architectures}(c), where we add a query translation component that feeds the query side of the bi-encoder architecture.

Query translation is much more computationally efficient than document translation, but has the disadvantages already discussed---queries may be more difficult to translate given that they may not be well-formed sentences.
However, this approach enables more rapid experimentation since the introduction of a new translation model does not require re-translation of the entire corpus.

One challenge of query translation is that we need a good mono-lingual retrieval model in $f$, which by definition is non-English.
While BM25 can provide a baseline (in the bag-of-words space of language $f$), effective learned retrieval models are more difficult to come by since less manually labeled data are available in non-English languages.

Our experiments consider both human and machine translations of the topics provided by the track organizers.
From each type of translation, we can create three types of queries: ``title'', ``description'', and ``both'' (similar to the document translation case above).
Thus, we have a total of six variations: \{human translation, machine translation\} $\times$ \{title, description, both\}. 
With these conditions, we experimented with two different retrieval models as well as pseudo-relevance feedback:

\paragraph{BM25}
Again, this traditional ``bag-of-words'' model remains a robust baseline.

\paragraph{SPLADE}
To build SPLADE models in non-English languages, we first need to start with a good pretrained language model for that language.
Thus, the models used here are first trained from scratch with the MLM+FLOPS loss~\cite{spladescratch} using a corpus concatenation of (i) the NeuCLIR corpus of the target language, (ii) the MS MARCO translations~\cite{mmarco} for the target language, and (iii) the Mr.\ TyDi~\cite{mrtydi} corpus of the target language (if available).
Finally, we fine-tuned on the target language version of MS MARCO, expecting to have similar zero-shot properties as similar experiments in English.
A separate model was created for each language.\footnote{SPLADE and pretrained models are made available at \url{https://huggingface.co/naver/modelname} with \texttt{modelname} = \texttt{neuclir22-\{pretrained,splade\}-\{fa,ru,zh\}}} 

\paragraph{Pseudo-relevance feedback}
As in the document translation case, we can apply pseudo-relevance feedback on top of either BM25 or SPLADE.
For the same reasons discussed above, Rocchio was chosen as the feedback method.

\subsection{Language-Independent Representations}

Starting from the bi-encoder design for mono-lingual retrieval shown in Figure~\ref{fig:architectures}(a), one might wonder if it were possible for the document and query encoders to generate some sort of language-independent semantic representation that would support direct relevance matching across languages.
With the advent of pretrained multilingual transformers, this is indeed possible.
For example, we can apply the document encoder to documents in $\mathcal{C}_f$ (in language $f$), and apply the query encoder to a query in $e$, and directly conduct relevance ranking on the representations.
Thus, we can perform cross-lingual retrieval without explicit query or document translation.
This is shown in Figure~\ref{fig:architectures}(d).

The most straightforward implementation of this approach is to train a DPR model~\cite{dpr}, but starting from a multilingual transformer backbone such as mBERT.
To our knowledge, \citet{asai-etal-2021-xor} was the first to propose such an approach.
More recently, \citet{Zhang_etal_arXiv2022} built on this basic design and introduced different approaches to exploit cross-lingual transfer by ``pre--fine-tuning'' on English data before further fine-tuning on the target languages using non-English data.
Although Zhang et al.~focused on mono-lingual retrieval in non-English languages, many of the lessons learned are applicable to the cross-lingual case as well.

Specifically, for this work, we pre--fine-tuned a multilingual DPR model initialized from an XLM-R~\cite{xlmr} backbone,\footnote{\url{https://huggingface.co/xlm-roberta-large}} dubbed xDPR.
The model was trained on the MS MARCO passage dataset~\cite{msmarco}, where both query and passage encoders share parameters.

With this trained model, we separately encoded the corpora in Persian, Russian, and Chinese.
It is perhaps worth emphasizing that the {\it same} model was used in all three cases.
For query encoding, we have a number of design choices.
Similar to document translation and query translation, we can use ``title'', ``description'', or ``both''.
Furthermore, we can encode queries either in $e$ or $f$.
In the first case, we are asking the encoder to {\it directly} project $e$ queries into the semantic space occupied by the $f$ documents.
In the second case, the query starts off in $f$, so the model is encoding a sequence in $f$ into the semantic space occupied by $f$ documents.
Thus, for each language, we arrive at a total of nine variations: \{original query, human translation, machine translation\} $\times$ \{title, description, both\}. 

Finally, on top of xDPR retrieved results, we can apply pseudo-relevance feedback using Rocchio's method, following the work of \citet{Li:2108.11044:2021,Li_etal_SIGIR2022}.
Thus, combined with~\citet{Yuqi}, we are able to implement Rocchio feedback consistently across both dense and sparse retrieval models.

\subsection{Reranking}
\label{section:methods:reranking}

In a standard multi-stage ranking architecture, the first-stage retriever generates a ranked list of candidates that are then processed by one or more reranking stages that aim to improve the ranking.
Reranking is also applicable in the cross-lingual case, but depending on the first-stage retriever, the candidate query/document pairs may either be in $e$ or  $f$.
In cases where both the queries and documents are in $e$, we can use a mono-lingual English reranker.

For the first-stage retrievers based on document translation, our experiments used monoT5, which is based on T5~\cite{Raffel_etal_JMLR2020}.
Reranking is performed {\it in English} with the following prompt:
\[ \texttt{Query: \{query\_text\} Document: \{doc\_text\}  Relevant:}\]
The model is asked to generate either the ``true'' or ``false'' token, from which we can extract the probability of relevance used to sort the candidates.
When the monoT5 model is fine-tuned on the MS MARCO passage dataset, it achieves state-of-the-art results on the TREC Deep Learning Tracks~\cite{trec_2019,trec_2020}, as well as impressive zero-shot effectiveness on BEIR~\cite{jeronymo2023inpars} and many other datasets~\cite{Roberts2019OverviewOT,zhang2020rapidly,rosa2021tune,rosa2022billions}.

For reranking first-stage retrievers based on query translation, we used a variant based on the multilingual version of T5 called mT5, which was pretrained on the multilingual mC4 dataset~\cite{xue2020mt5}; otherwise, we use the same reranking approach.
To fine-tune mT5 for reranking, we employed a similar strategy as \citet{mmarco} using mMARCO, the multilingual version of the MS MARCO dataset.
For our experiments, we used the XXL model with 13B parameters.

\subsection{Fusion}

Researchers have known for many decades that fusion techniques, which combine evidence from multiple individual runs, can improve effectiveness~\cite{Bartell_etal_SIGIR1994,Vogt_Cottrell_1999}.
Fusion works particularly well when the individual runs are based on different underlying techniques, such as in the case of dense vs.\ sparse retrieval models~\cite{Gao_etal_ECIR2021,Ma_etal_arXiv2021_DPR}.
Given that our first-stage retrievers are all based on very different approaches, we would expect fusion to yield substantial boosts in effectiveness, although this does not appear to be borne out experimentally.

Within a multi-stage architecture, there arises a natural question of {\it when} fusion should be performed.
One possible approach is to independently rerank the output of each first-stage retriever, and then fuse those results; we call this late fusion.
Another possible approach is to first fuse the output of the first-stage retrievers, and then rerank the combined results; we call this early fusion.
The effectiveness difference between the two approaches is an empirical question, but late fusion is more computationally intensive because it requires more reranking.

\section{Implementation Details}
\label{section:implementation}

All the first-stage and fusion retrieval conditions described in this paper are implemented in Anserini~\cite{anserini} and Pyserini~\cite{lin2021pyserini}.
Anserini is a Java-based toolkit built around the open-source Lucene search library to support reproducible information retrieval research.
Pyserini provides a Python interface to Anserini and further augments its capabilities by including support for dense retrieval models.
Together, the toolkits are widely adopted by researchers in the IR and NLP communities.

For document translation using BM25, our implementation uses Lucene's default analyzer for English, which performs tokenization, stemming, etc.
Retrieval is performed with Pyserini's default BM25 parameters ($k_1=0.9$, $b=0.4$).
For query translation, note that since we are indexing non-English text, analyzers in $f$ are required.
Fortunately, Lucene already has analyzers implemented for all three languages, which we used out of the box.
The same BM25 parameters were used.

All SPLADE models were implemented in Lucene using the standard ``fake documents'' trick~\cite{Mackenzie_etal_arXiv2021}.
Token weights were used to generate synthetic documents where the token was repeated a number of times equal to its weight (after quantizing into integers).
For example, if ``car'' receives a weight of ten from the encoder, we simply repeat the token ten times.
These fake documents are then indexed with Anserini as usual, where the weight is stored in the term frequency position of the postings in the inverted index.
Top-$k$ retrieval is implemented by using a ``sum of term frequency'' scoring function in Lucene, which produces exactly the same output as ranking by the inner product between query and document vectors.
Anserini provides the appropriate abstractions that hide all these implementation details.

Support for dense retrieval is provided in Pyserini with the Faiss toolkit~\cite{faiss}; all xDPR runs were conducted with flat indexes.
For both BM25 and SPLADE models, Anserini exposes the appropriate bindings for performing retrieval in Python, and Pyserini provides appropriate interfaces that abstract over and unify retrieval using dense and sparse models (i.e., they are merely parametric variations in the command-line arguments).
Pyserini additionally provides implementations of reciprocal rank fusion, and thus the entire infrastructure makes mixing-and-matching different experimental conditions quite easy.

\begin{table*}[t]
\begin{tabular}{llllllllllll}
\toprule
\multicolumn{2}{l}{\bf nDCG@20} & & \multicolumn{3}{c}{\textbf{Persian}} & \multicolumn{3}{c}{\textbf{Russian}} & \multicolumn{3}{c}{\textbf{Chinese}}\\
\cmidrule(lr){4-6} \cmidrule(lr){7-9} \cmidrule(lr){10-12}
 &  & PRF  & title & desc & both & title & desc & both & title & desc & both \\
\midrule
\multicolumn{6}{l}{{\bf document translation --- BM25}} \\
(1a) & official Sockeye translation & \xmark  & 0.3665 & 0.2889 & 0.3670 & 0.3693 & 0.2060 & 0.3080 & 0.3705 & 0.3070 & 0.3723 \\
(1b) & official Sockeye translation & \cmark & 0.3532 & 0.3127 & 0.3720 & 0.3589 & 0.2627 & 0.3188 & 0.3802 & 0.3206 & 0.3806 \\
\midrule
\multicolumn{6}{l}{{\bf document translation --- SPLADE}} \\
(2a) & official Sockeye translation & \xmark & 0.4627 & 0.4618 & 0.4802 & 0.4865 & 0.4193 & 0.4573 & 0.4233 & 0.4299 & 0.4236 \\
(2b) & official Sockeye translation & \cmark   & 0.4438 & 0.4675 & 0.4645 & 0.4836 & 0.4243 & 0.4604 & 0.4204 & 0.4142 & 0.4206 \\
\midrule
\multicolumn{6}{l}{{\bf query translation --- BM25}} \\
(3a) & human translation (HT) & \xmark  & 0.3428 & 0.2843 & 0.3429 & 0.3668 & 0.3138 & 0.3665 & 0.2478 & 0.2068 & 0.2572 \\
(3b) & machine translation (MT) & \xmark   & 0.3331 & 0.2974  & 0.3700  & 0.3564 & 0.2972 & 0.3605 & 0.1830 & 0.1498 & 0.1754 \\
(3c) & human translation (HT) & \cmark   & 0.3356  & 0.2885 &  0.3408 & 0.3572 & 0.3366 & 0.3630 & 0.2544 & 0.1985 & 0.2734 \\
(3d) & machine translation (MT) & \cmark  & 0.3374 & 0.3300 & 0.3612 & 0.3426 & 0.3257 & 0.3764 & 0.1861 & 0.1464 & 0.1785 \\
\midrule
\multicolumn{6}{l}{{\bf query translation --- SPLADE}} \\
(4a) & human translation (HT) & \xmark  & 0.4301 & 0.4413 & 0.4788 & 0.4594 & 0.3922 & 0.4214 & 0.3110 & 0.2935 & 0.3143 \\
(4b) & machine translation (MT) & \xmark   & 0.4437 & 0.4300  & 0.4728 & 0.4452 & 0.3792 & 0.4156 & 0.2843 & 0.2527 & 0.2929 \\
(4c) & human translation (HT) & \cmark   &  0.4348 & 0.4232 & 0.4146 & 0.4322 & 0.4133 & 0.4316 & 0.3198 & 0.2926 & 0.3077 \\
(4d) & machine translation (MT) & \cmark   & 0.4193 & 0.4121 & 0.4444 & 0.4337 & 0.3965 & 0.4075 & 0.2920 & 0.2562 & 0.3029 \\
\midrule
\multicolumn{6}{l}{{\bf language-independent representations --- xDPR}} \\
(5a) & $\langle$d: original corpus, q: English$\rangle$ & \xmark & 0.1522 & 0.1847 & 0.1804 & 0.2967 & 0.2913 & 0.2866 & 0.2200 & 0.2192 & 0.2185 \\
(5b)  &  $\langle$d: original corpus, q: HT$\rangle$  &  \xmark  &  0.2776 & 0.2900 & 0.2953 & 0.3350 & 0.3276 & 0.3307 & 0.3197 & 0.3129 & 0.3035\\
(5c)  &  $\langle$d: original corpus, q: MT$\rangle$  &  \xmark  &  0.2721 & 0.2968 & 0.3055 & 0.3619 & 0.3348 & 0.3542 & 0.3025 & 0.2785 & 0.3013 \\
(5d)  &  $\langle$d: original corpus, q: English$\rangle$  &  \cmark  &  0.1694 & 0.1996 & 0.1993 & 0.3116 & 0.3085 & 0.3045 & 0.2442 & 0.2343 & 0.2312 \\
(5e)  &  $\langle$d: original corpus, q: HT$\rangle$  &  \cmark  &  0.3083 & 0.2988 & 0.3197 & 0.3349 & 0.3544 & 0.3578 & 0.3376 & 0.3463 & 0.3380 \\
(5f)  &  $\langle$d: original corpus, q: MT$\rangle$  &  \cmark  &  0.3136 & 0.3012 & 0.3181 & 0.3727 & 0.3690 & 0.3793 & 0.3268 & 0.3041 & 0.3345 \\
\bottomrule
\end{tabular}
\vspace{0.2cm}
\caption{Main results table reporting nDCG@20 for various first-stage retrievers.}
\label{table:first-stage-ndcg}
\end{table*}

\begin{table*}[t]
\begin{tabular}{llllllllllll}
\toprule
\multicolumn{2}{l}{\bf Recall@1000} & & \multicolumn{3}{c}{\textbf{Persian}} & \multicolumn{3}{c}{\textbf{Russian}} & \multicolumn{3}{c}{\textbf{Chinese}}\\
\cmidrule(lr){4-6} \cmidrule(lr){7-9} \cmidrule(lr){10-12}
 &  & PRF  & title & desc & both & title & desc & both & title & desc & both \\
\midrule
\multicolumn{6}{l}{{\bf document translation --- BM25}} \\
(1a) & official Sockeye translation & \xmark & 0.7335 & 0.6319 & 0.7652 & 0.7409 & 0.5780 & 0.7255 & 0.7567 & 0.6639 & 0.7567 \\
(1b) & official Sockeye translation & \cmark & 0.8111 & 0.7638 & 0.8248 & 0.7908 & 0.6780 & 0.7798 & 0.8129 & 0.7404 & 0.8011 \\
\midrule
\multicolumn{6}{l}{{\bf document translation --- SPLADE}} \\
(2a) & official Sockeye translation & \xmark & 0.8478 & 0.8796 & 0.8860 & 0.8538 & 0.8376 & 0.8513 & 0.7997 & 0.7597 & 0.7922 \\
(2b) & official Sockeye translation & \cmark & 0.8592 & 0.8735 & 0.8703 & 0.8686 & 0.8238 & 0.8544 & 0.8038 & 0.7623 & 0.8067 \\
\midrule
\multicolumn{6}{l}{{\bf query translation --- BM25}} \\
(3a) & human translation (HT)   & \xmark & 0.7128 & 0.7027 & 0.7373 & 0.7125 & 0.6655 & 0.7421 & 0.4759 & 0.4577 & 0.4940 \\
(3b) & machine translation (MT) & \xmark & 0.7254 & 0.6815 & 0.7424 & 0.7332 & 0.6210 & 0.7373 & 0.3829 & 0.2989 & 0.4028 \\
(3c) & human translation (HT)   & \cmark & 0.7691 & 0.7520 & 0.8092 & 0.7381 & 0.7276 & 0.7770 & 0.5230 & 0.5113 & 0.5327 \\
(3d) & machine translation (MT) & \cmark & 0.7672 & 0.7033 & 0.7829 & 0.7439 & 0.7136 & 0.7959 & 0.4361 & 0.3748 & 0.4341 \\
\midrule
\multicolumn{6}{l}{{\bf query translation --- SPLADE}} \\
(4a) & human translation (HT)   & \xmark & 0.7652 & 0.8173 & 0.8239 & 0.7739 & 0.7200 & 0.7612 & 0.6803 & 0.6602 & 0.6551 \\
(4b) & machine translation (MT) & \xmark & 0.8045 & 0.8172 & 0.8437 & 0.7725 & 0.7150 & 0.7669 & 0.6424 & 0.5919 & 0.6312 \\
(4c) & human translation (HT)   & \cmark & 0.7897 & 0.8175 & 0.8245 & 0.7946 & 0.7209 & 0.7776 & 0.7100 & 0.7205 & 0.7029 \\
(4d) & machine translation (MT) & \cmark & 0.8099 & 0.8117 & 0.8350 & 0.7918 & 0.7090 & 0.7590 & 0.6861 & 0.6096 & 0.6535 \\
\midrule
\multicolumn{6}{l}{{\bf language-independent representations --- xDPR}} \\
(5a) & $\langle$d: original corpus, q: English$\rangle$ & \xmark & 0.4910 & 0.5445 & 0.5393 & 0.5704 & 0.5627 & 0.5834 & 0.4161 & 0.4359 & 0.4386\\
(5b)  &  $\langle$d: original corpus, q: HT$\rangle$  &  \xmark  &  0.6288 & 0.6780 & 0.7088 & 0.6196 & 0.5825 & 0.6368 & 0.5773 & 0.5841 & 0.6031  \\
(5c)  &  $\langle$d: original corpus, q: MT$\rangle$  &  \xmark  &  0.6333 & 0.6453 & 0.6850 & 0.6285 & 0.5649 & 0.6300 & 0.5420 & 0.5382 & 0.5873 \\
(5d)  &  $\langle$d: original corpus, q: English$\rangle$  &  \cmark  &  0.4702 & 0.4981 & 0.5347 & 0.6251 & 0.5971 & 0.6212 & 0.4330 & 0.4714 & 0.4593 \\
(5e)  &  $\langle$d: original corpus, q: HT$\rangle$  &  \cmark  &  0.6409 & 0.6612 & 0.7212 & 0.6541 & 0.5915 & 0.6346 & 0.6088 & 0.5939 & 0.6310\\
(5f)  &  $\langle$d: original corpus, q: MT$\rangle$   &  \cmark  &  0.6686 & 0.6516 & 0.7071 & 0.6784 & 0.6032 & 0.6475 & 0.5744 & 0.5375 & 0.6109\\
\bottomrule
\end{tabular}
\vspace{0.2cm}
\caption{Main results table reporting recall@1000 for various first-stage retrievers.}
\label{table:first-stage-recall}
\end{table*}

\section{Results}

Our results are organized into following progression:\ first-stage retrievers, reranking, and fusion.
We report retrieval effectiveness in terms of nDCG@20, the official metric of the NeuCLIR evaluation, and recall at a cutoff of 1000 hits (recall@1000), which quantifies the effectiveness upper bound of reranking.
The organizers also measured mean average precision (MAP) as a supplemental metric;
we followed this procedure as well.
Overall, the findings from nDCG@20 and MAP were consistent, and so for brevity we omit the MAP results in our presentation.

In Section~\ref{section:clir}, we describe a vast design space for first-stage variants that can feed many reranking and fusion approaches.
It is not practical to exhaustively examine all possible combinations, and thus our experiments were guided by progressive culling of ``uninteresting'' settings, as we'll describe.

Finally, a word on significance testing:\
We are of course cognizant of its importance, but we are equally aware of the dangers of multiple hypothesis testing.
Due to the large number of conditions we examine, a standard technique such as the Bonferroni correction is likely too conservative to detect significant differences, especially given the relatively small topic size of NeuCLIR.
For most of our experiments, we did {\it not} perform significance testing and instead focused on general trends that are apparent from our large numbers of experimental conditions.
We applied significance testing more judiciously, to answer targeted research questions.
To be clear, the results we report are the only tests we conducted---that is, we did not cherry-pick the most promising results.
In all cases, we used paired $t$-tests ($p\leq0.05$) with the Bonferroni correction.

\subsection{First-Stage Retrievers}

We begin by examining the output of individual first-stage retrievers.
Tables~\ref{table:first-stage-ndcg} and~\ref{table:first-stage-recall} present results in terms of nDCG@20 and recall@1000, respectively.
Each block of rows is organized by the general approach.
The columns show metrics grouped by language, and within each block, we report the results of using queries comprised of the ``title'' field, the ``description'' field, and both.

\paragraph{Document translation}
Recall that in the document translation condition, we are indexing the machine-translated documents provided by the NeuCLIR organizers, which are in English.
The BM25 conditions in rows (1ab) and the SPLADE conditions in rows (2ab) differ only in the retrieval model applied to the translated corpus.
For BM25, we see that ``title'' and ``both'' query conditions yield about the same effectiveness (both metrics) on Persian and Chinese, but ``both'' is worse on Russian.
For all languages, it appears that ``description'' queries perform worse.
For SPLADE, interestingly, for Persian and Chinese, there does not appear to be much of an effectiveness gap between the three types of queries for both metrics.
This is likely because the retrieval model includes query expansion, and so the benefits from having richer descriptions of the information need diminish.

The comparisons between (a) vs.\ (b) rows highlight the impact of pseudo-relevance feedback.
We see that, at best, PRF yields a small improvement for BM25 in terms of nDCG@20, and for SPLADE, PRF actually decreases effectiveness.
However, looking at the recall figures in Table~\ref{table:first-stage-recall}, it does appear that PRF boosts recall.
This behavior is expected, as PRF is primarily a recall-enhancing device.

\paragraph{Query translation}
With BM25, shown in rows (3a)--(3d), we see that ``title'' and ``both'' conditions are generally on par for Russian and Chinese for both metrics.
For SPLADE, shown in rows (4a)--(4d), there does not appear to be a consistent finding:\ in some cases, ``both'' beats ``title'', and the opposite in other cases.
However, it does appear that ``description'' alone is generally less effective in terms of nDCG@20.

With query translation, there is a natural comparison between human translations and machine translations.
In rows (3) and (4), these are the (a) and (c) conditions versus the (b) and (d) conditions.
It does not appear that for Persian and Russian, machine-translated queries are consistently less effective than human translations, for both BM25 and SPLADE.
In some cases, we actually observe machine-translated queries outperforming their human-translation counterparts.
For BM25, note that since the queries are bags of words, the fluency of the translations is not important, so long as the correct content terms are present.
For SPLADE, the model appears to be robust to possibly disfluent translations.
In Chinese, however, there does seem to be a noticeable gap between human and machine translations, with the human translations generally yielding better results.

Finally, consistent with the document translation case, pseudo-relevance feedback does not appear to improve nDCG@20, but does improve recall.
Once again, this is expected.

\paragraph{Language-Independent Representations}
The final blocks in Tables~\ref{table:first-stage-ndcg} and~\ref{table:first-stage-recall} show the effectiveness of xDPR. 
Recall our experimental design:\ on the document end, the original corpus in $f$ is encoded with the model.
On the query end, there are three options:\ directly encode the English query, encode the human-translated (HT) query, or encode the machine-translated (MT) query.
These are shown in rows (5a), (5b), and (5c), respectively.
We see quite a big difference in effectiveness between row (5a) and row (5b), which indicates that there is a big loss in trying to encode queries in $e$ directly into the semantic space occupied by documents in $f$, compared to encoding queries in $f$.
Clearly, the model is not able to adequately encode text with the same meaning in different languages (the query translations) into the same semantic space.
Regardless of configuration, the dense retrieval models appear to be far less effective than the BM25 and SPLADE models, for both translation types, across both metrics.
However, we see that pseudo-relevance feedback does appear to increase effectiveness, which is consistent with previous work~\cite{Li:2108.11044:2021,Li_etal_SIGIR2022} on vector PRF.

\subsection{Reranking}

\begin{table*}[t]
\begin{tabular}{llllllllllllll}
\toprule
\multicolumn{2}{l}{\bf nDCG@20} & \multicolumn{2}{c}{\textbf{Persian}} & \multicolumn{2}{c}{\textbf{Russian}} & \multicolumn{2}{c}{\textbf{Chinese}}\\
\cmidrule(lr){3-4} \cmidrule(lr){5-6} \cmidrule(lr){7-8}
 &  & 1st & rerank & 1st & rerank & 1st & rerank \\
\midrule
\multicolumn{6}{l}{{\bf document translation --- BM25}} \\
(1a) & official Sockeye translation & (0.3670, 0.7652) & 0.5350 & (0.3080, 0.7255) & 0.5662 & (0.3723, 0.7567) & 0.4955 \\
\midrule
\multicolumn{6}{l}{{\bf document translation --- SPLADE}} \\
(2a) & official Sockeye translation & (0.4802, 0.8860) & 0.5545 & (0.4573, 0.8513) & 0.5714 & (0.4236, 0.7922) & 0.5026 \\
\midrule
\multicolumn{6}{l}{{\bf query translation --- BM25}} \\
(3a) & human translation (HT) & (0.3429, 0.7373) & 0.5346 & (0.3665, 0.7421) & 0.5745 & (0.2572, 0.4940) & 0.4300  \\
(3b) & machine translation (MT) & (0.3700, 0.7424) & 0.5551 & (0.3605, 0.7373) & 0.5742 & (0.1754, 0.4028) & 0.3831 \\
\midrule
\multicolumn{6}{l}{{\bf query translation --- SPLADE}} \\
(4a) & human translation (HT) & (0.4788, 0.8239) & 0.5722 & (0.4214, 0.7612) & 0.5823 & (0.3143, 0.6551) & 0.4980  \\
(4b) & machine translation (MT) & (0.4728, 0.8437) & 0.5932 & (0.4156, 0.7669) & 0.5767 & (0.2929, 0.6312) & 0.5132 \\
\midrule
\multicolumn{6}{l}{{\bf language-independent representations --- xDPR}} \\
(5a) & $\langle$d: original corpus, q: English$\rangle$ & (0.1804, 0.5393) & 0.4630 & (0.2866, 0.5834) & 0.5305 & (0.2185, 0.4386) & 0.4440\\
(5b) & $\langle$d: original corpus, q: HT$\rangle$ & (0.2953, 0.7088) & 0.5614 & (0.3307, 0.6368) & 0.5617 & (0.3035, 0.6031) & 0.5008\\
(5c) & $\langle$d: original corpus, q: MT$\rangle$ & (0.3055, 0.6850) & 0.5644 & (0.3542, 0.6300) & 0.5337 & (0.3013, 0.5873) & 0.5087\\
\bottomrule
\end{tabular}
\vspace{0.2cm}
\caption{Results of reranking various first-stage retrievers (nDCG@20).
Under the column ``1st'' we repeat the (nDCG@20, Recall@1000) metrics from the first-stage retriever for convenience. In all cases we used both titles and descriptions as queries in first-stage retrieval (with no pseudo-relevance feedback) and reranking.}
\label{table:reranking-single-runs}
\end{table*}
\begin{table*}[t]
\begin{tabular}{lllllllll}
\toprule
 & &  \multicolumn{3}{c}{\textbf{nDCG@20}} & \multicolumn{3}{c}{\textbf{Recall@1000}} \\
 \cmidrule(lr){3-5} \cmidrule(lr){6-8}
 & & \multicolumn{1}{c}{\textbf{Persian}} & \multicolumn{1}{c}{\textbf{Russian}} & \multicolumn{1}{c}{\textbf{Chinese}} & \multicolumn{1}{c}{\textbf{Persian}} & \multicolumn{1}{c}{\textbf{Russian}} & \multicolumn{1}{c}{\textbf{Chinese}} \\
\midrule
(1a) & DT--BM25    & 0.3670 & 0.3080 & 0.3723 & 0.7652 & 0.7255 & 0.7567 \\
(2a) & DT--SPLADE  & 0.4802 & 0.4573 & 0.4236 & 0.8860 & 0.8513 & 0.7922 \\
(3b) & QT--BM25    & 0.3700 & 0.3605 & 0.1754 & 0.7424 & 0.7373 & 0.4028 \\
(4b) & QT--SPLADE  & 0.4728 & 0.4156 & 0.2929 & 0.8437 & 0.7669 & 0.6312 \\
(5a) & dense--e    & 0.1804 & 0.2866 & 0.2185 & 0.5393 & 0.5834 & 0.4386 \\
(5c) & dense--f    & 0.3055 & 0.3542 & 0.3013 & 0.6850 & 0.6300 & 0.5873 \\
\midrule
(6a) & RRF(1a, 2a): DT--BM25, DT--SPLADE & \red{0.4462} & \red{0.4180} & \red{0.4189} & 0.8936 & 0.8670 & 0.8536\\
(6b) & RRF(3b, 4b): QT--BM25, QT--SPLADE  & \textcolor{red}{0.4610} & 0.4598 & 0.2981 & 0.8703 & 0.8368 & 0.6692 \\
(6c) & RRF(1a, 3b): DT--BM25, QT--BM25 & 0.3795 & 0.3635 & \textcolor{red}{0.2736} & 0.7901 & 0.7686 & \red{0.7366}  \\
(6d) & RRF(2a, 4b): DT--SPLADE, QT--SPLADE & 0.5165 & 0.4921 & \textcolor{red}{0.4178} & 0.9009 & \red{0.8508} & 0.7938 \\
(6e) & RRF(1a, 2a, 3b, 4b): DT, QT & 0.4897 & 0.4857 & 0.4397 & 0.9285$^\dagger$ & 0.8880 & 0.8637$^\dagger$ \\
(6f) & RRF(5a, 5c): dense & \textcolor{red}{0.2640} & \textcolor{red}{0.3469} & \textcolor{red}{0.2731} & \red{0.6814} & 0.6493 & \red{0.5693} \\
(6g) & RRF(1a, 2a, 3b, 4b, 5a, 5c): DT, QT, dense & 0.4926 & 0.5142$^\dagger$ & 0.4541 & 0.9291$^\dagger$ & 0.8818 & 0.8704$^\dagger$ \\
\bottomrule
\end{tabular}
\vspace{0.2cm}
\caption{Results of different fusion combinations. Scores of individual first-stage retrievers are repeated for convenience. In all cases we used both titles and descriptions as queries, with no pseudo-relevance feedback. Red shows cases where fusion performed worse than the best single input run. $^\dagger$ represents a significant improvement over (2a).}
\label{table:fusion}
\end{table*}

In the previous section, we examined first-stage retrieval settings for 18 $\times$ 3 $=$ 54 different conditions, for each language.
It is impractical to report reranking results for every single condition, and thus we made a few choices to focus our attention:
We considered only conditions that take advantage of both title and description fields, which appear to be more robust than title-only queries.
We also focused on runs without PRF, since PRF represents additional computational costs (both latency and index size).

For each language, this reduces the number of first-stage retrievers under consideration to nine.
We applied reranking on these runs, including the title and description fields in the input template to the reranking models.
We informally, but not exhaustively, examined other conditions, but they did not appear to alter our overall findings.
For example, we tried reranking the first-stage retrieval results with pseudo-relevance feedback, but the results were not noticeably better (even though they exhibited higher recall).

Reranking results are shown in Table~\ref{table:reranking-single-runs}.
Under the effectiveness of the first-stage retriever (``1st'' columns), we report (nDCG@20, recall@1000):\ the first quantifies candidate ranking quality and the second quantifies the upper bound effectiveness of a reranker.
We see that reranking improves effectiveness by large margins, but this is expected as the effectiveness of cross-encoders in various settings is well known (see Section~\ref{section:methods:reranking}).

One interesting observation, however, is that reranking reduces the effectiveness gap between the best and worst first-stage retrievers.
For example, starting with BM25, which is clearly less effective than SPLADE, the reranker is able to ``make up'' for the lower quality candidates, such that the end-to-end effectiveness is relatively close to reranking SPLADE results (at least in terms of nDCG).
In fact, in some cases, reranking xDPR results yields scores that are even higher than reranking BM25 results.
While ``coupling effects'' between the first-stage retriever and reranker have been previously noted in the literature~\cite{Gao_etal_ECIR2021,Pradeep_etal_ECIR2022}, this finding affirms the need for further explorations.

\subsection{Fusion}

With fusion, the design space of possible combinations is immense and impractical to exhaustively explore.
To provide continuity, we focus only on the first-stage retrievers in the reranking experiments.
In the space of fusion techniques, we settled on reciprocal rank fusion, which is a simple, effective, and robust approach~\cite{Cormack_etal_SIGIR2009}.

\begin{table*}[t]
\begin{tabular}{lllllllllll}
\toprule
 & & \multicolumn{3}{c}{\textbf{Persian}} & \multicolumn{3}{c}{\textbf{Russian}} & \multicolumn{3}{c}{\textbf{Chinese}} \\
 \cmidrule(lr){3-5} \cmidrule(lr){6-8} \cmidrule(lr){9-11}
& & 1st & early & late & 1st & early & late & 1st & early & late \\
\midrule
(4a) & QT--SPLADE = best single & 0.4728 & 0.5932 & & 0.4156 & 0.5767 & & 0.2929 & 0.5132 \\
\midrule
(6c) & RRF(1a, 3b): DT--BM25, QT--BM25 & 0.3795 & 0.5869 & 0.5723 &0.3635 & 0.5788 & 0.5890 & 0.2736 & 0.5257$^\dagger$ & 0.4150 \\
(6d) & RRF(2a, 4b): DT--SPLADE, QT--SPLADE & 0.5165 & 0.5823 & 0.6122 & 0.4921 & 0.5729 & 0.5915 & 0.4178 & 0.5379 & 0.5272 \\
(6e) & RRF(1a, 2a, 3b, 4b): DT, QT & 0.4897 & 0.5901 & 0.5911 & 0.4857 & 0.5728 & 0.5853 & 0.4397 & 0.5394 & 0.5058 \\
(6f) & RRF(5a, 5c): dense & 0.2640 & 0.5621$^\dagger$ & 0.4573 & 0.3469 & 0.5438 & 0.5162 & 0.2731 & 0.5077$^\dagger$ & 0.4470\\
(6g) & RRF(1a, 2a, 3b, 4b, 5a, 5c): DT, QT, dense & 0.4926 & 0.5893 & 0.5626 & 0.5142 & 0.5676 & 0.5840 & 0.4541 & 0.5340 & 0.5295 \\
\bottomrule
\end{tabular}
\vspace{0.25cm}
\caption{Comparisons between early and late fusion. $^\dagger$ represents a significant improvement over late fusion.}
\label{table:e2e}
\end{table*}

With these considerations, we experimented with the following fusion conditions in Table~\ref{table:fusion}:\
(6a) document translation combining BM25 and SPLADE;
(6b) query translation combining BM25 and SPLADE;
(6c) combining document and query translation with BM25;
(6d) combining SPLADE document and query translation;
(6e) combining all lexical approaches;
(6f) combining both dense approaches;
(6g) combining everything.
The top block of Table~\ref{table:fusion} repeats the effectiveness of the first-stage retrievers for convenience.
In the bottom block of the table, cases in which the fusion results are worse than the best input are shown in red.
In these cases, fusion provides no value over just selecting the best individual run.

From these results, it appears that for Persian and Russian, the best effectiveness can be achieved by fusing both document translation and query translation SPLADE models, row (6d), although for Chinese, the same fusion is a bit worse than just document translation SPLADE.
Fusing all the lexical runs, row (6e), is a bit worse than fusing just SPLADE runs in Persian and Russian, but it improves Chinese.
Finally, incorporating evidence from the language-independent dense retrieval techniques appears to provide value over simply fusing the lexical results, as we see comparing (6g) and (6e).
This is surprising given that by themselves, the dense retrieval runs are quite poor.

Overall, we were somewhat surprised by the finding that fusion did not improve effectiveness as robustly as we had hoped.
In Table~\ref{table:fusion}, the figures in red represent all the cases in which fusion actually {\it hurt} effectiveness, i.e., fusion performed worse than the best single input run.
We attribute this finding to the large differences in effectiveness between the runs, in that RRF does not work as well if one of the fusion inputs is much better than the others.

To more rigorously test this observation, we performed significance testing comparing the document translation SPLADE model, row (2a) in Table~\ref{table:fusion}, against fusion of SPLADE models, row (6d), fusion of all lexical models, row (6e), and fusion of all lexical and dense models, row (6g).
These comparisons answer the following questions, starting from the single best first-stage retriever:\ Does SPLADE fusion provide any additional value?
What about BM25?
Dense retrieval?

The conclusion, reported in Table~\ref{table:fusion} with the symbol $^\dagger$, is that most of the fusion combinations are not statistically significantly better than document translation with SPLADE, the single best first-stage retriever.
For nDCG@20, the largest ensemble is significantly better than DT--SPLADE only on Russian; for recall@1000 we see more significant improvements, but only on Persian and Chinese.
Notably, combining evidence from both document and query translation with SPLADE, row (6d), is not significantly better than DT--SPLADE alone.

In our final set of experiments, we compared the effectiveness between early and late fusion for a subset of the conditions in Table~\ref{table:fusion}.
These results are reported in Table~\ref{table:e2e}.
In this case, we use QT--SPLADE as the point of comparison, which appears to provide the best single-stage retriever and reranking combination.
For Persian, late fusion appears to be either about the same or slightly better, with the exception of (6f); this appears to be the case for Russian also, although the late fusion margin of improvement seems to be smaller.
Chinese results are a bit more mixed, with early beating late in some cases.
To more rigorously compare early vs.\ late fusion, we performed significance tests comparing all pairs.
Only a few of these differences are significant, and they only happen for cases where early fusion is better than late fusion.
Two of the three cases, however, occurred for the dense models, which are less effective to begin with.
Overall, these experiments are inconclusive with respect to the question of which fusion strategy is better.

To provide additional context, the best runs from the NeuCLIR 2022 evaluation were from members of our group, but were generated under the time pressure of deadlines and thus it was not possible to carefully consider all configurations as we did in Table~\ref{table:e2e}.
The best runs were (nDCG@20 scores):\
(i) Persian:\ \texttt{p2.fa.rerank}, 0.588;
(ii) Russian:\ \texttt{p3.ru.mono}, 0.567;
(iii) Chinese:\ \texttt{p2.zh.rerank}, 0.516.
Comparing those runs to the best conditions reported here, we verify that just by carefully studying the various effects of different system components, improvements are possible across all languages, achieving new state-of-the-art effectiveness with (i) Persian:\ 6d late-fusion 0.612 ($+$0.024);
(ii) Russian:\ 6d late-fusion 0.592 ($+$0.025);
(iii) Chinese:\ 6e early-fusion 0.539 ($+$0.023). 

\section{Conclusions}

The NeuCLIR evaluation at TREC 2022 represents a ``revival'' of interest in the cross-lingual information retrieval challenge in the ``neural era''.
As a high-level summary, this work captures a collaboration between two teams that submitted the most effective runs and a research group that is experienced in building retrieval toolkits to support research.
Together, we take a more principled approach to the panoply of methods that were deployed in the evaluation and provide an organizing conceptual framework based on mono-lingual retrieval.

What are the high-level takeaways?
It appears that query translation and document translation, general approaches dating back decades, adapt well to the neural age.
In particular, SPLADE appears to be highly effective, demonstrating the promise of sparse learned representations.
Although language-independent representations do not appear to be as effective as either query or document translation, we have only begun to scratch the surface of this class of techniques, as xDPR can only be considered a baseline.
But if one considers that the {\it same} xDPR model works for all three languages, we can see tremendous potential.

Across the NLP and IR communities, we have only begun to explore the application of large pretrained transformer models.
We find future prospects very exciting, and believe that our conceptual framework, experimental results, and software infrastructure offer a solid foundation for further exploration.

\section*{Acknowledgements}

This research was supported in part by the Natural Sciences and Engineering Research Council (NSERC) of Canada.

\bibliographystyle{ACM-Reference-Format}
\bibliography{ref}

\end{document}